%% file: ms.tex
\documentclass[12pt,preprint,psfig]{aastex}
\usepackage{color}
\shorttitle{The Chandra HETGS view of 4U 1624$-$490}
\shortauthors{Xiang et al.}

\def\1624{4U~1624$-$490\,}
\def\grs1915{GRS~1915+105\,}

\def\cyg{Cygnus~X-2}
\def\chandra{{\it Chandra\,}}
\def\xmm{{\it XMM-Newton\,}}
\def\beppo{{\it BeppoSAX\,}}
\def\asca{{\it ASCA\,}}

\def\exosat{{\it EXOSAT\,}}
\def\einstein{{\it Einstein\,}}
\def\xte{{\it RXTE\,}}
\def\ginga{{\it Ginga\,}}

\def\feI{Fe {\sc xxvi}\,}
\def\feII{Fe {\sc xxv}\,}

\def\hot{{\it ``hot''}\,}
\def\warm{{\it ``warm''}\,}

\begin{document}

\title{The accretion disk corona and disk atmosphere of \1624 as viewed 
  by the Chandra-HETGS}

\author{Jingen Xiang\altaffilmark{1}, Julia C. Lee\altaffilmark{1}}
\affil{\altaffilmark{1}Harvard University Department of Astronomy
(a part of the Harvard-Smithsonian Center for Astrophysics), 
60 Garden Street, Cambridge, MA 02138}
\email{jxiang@cfa.harvard.edu; jclee@cfa.harvard.edu}
\and
\author{Michael A. Nowak\altaffilmark{2}, J\"orn Wilms \altaffilmark{3}, Norbert S. Schulz\altaffilmark{2}}
\affil{\altaffilmark{2}Massachusetts Institute of Technology, Kavli Center, 
77 Massachusetts Ave.\ NE80, Cambridge, MA 02139}
\affil{\altaffilmark{3}Dr.\ Karl-Remeis-Observatory and Erlangen
  Centre for Astroparticle Physics, University of Erlangen-Nuremberg,
  Sternwartstr.~7, 96049 Bamberg, Germany}

\begin{abstract}
We present a detailed spectral study (photoionization modelling and
variability) of the ``Big Dipper'' \1624 based on a \chandra-High
Energy Transmission Gratings Spectrometer (HETGS) observation over the
$\sim76$~ks binary orbit of \1624. While the continuum spectrum can be
modeled using a blackbody plus power-law, a slightly
  better fit is obtained using a single $\Gamma=2.25$ power-law
partially (71\%) covered by a local absorber of column density $N_{\rm
  H,\ Local}=8.1_{-0.6}^{+0.7}\times 10^{22}\, \rm cm^{-2}$.
The data show a possible quasi-sinusoidal modulation
  with period $43_{-9}^{+13}$~ks that might be due to changes in local
  obscuration. Photoionization modeling with the {\sc xstar} code
and variability studies of the observed strong \ion{Fe}{25} and
\ion{Fe}{26} absorption lines point to a two-temperature plasma for
their origin: a highly ionized component of ionization parameter
$\xi_{\rm hot} \approx 10^{4.3}\,{\rm ergs\,cm\,s^{-1}}$ ($T\sim
3.0\times 10^{6}$~K) associated with an extended accretion disk corona
of radius $R \sim3\times10^{10}$~cm, and a less ionized more variable
component of $\xi \approx 10^{3.4}\,{\rm ergs\,cm\,s^{-1}}$ ($T\sim
1.0\times 10^{6}$~K) and $\rm \xi \approx 10^{3.1}\,
ergs\,cm\,s^{-1}$ ($T\sim 0.9\times 10^{6}$~K) coincident
with the accretion disk rim. We use this, with the observed
\ion{Fe}{25} and \ion{Fe}{26} absorption line variations (in
wavelength, strength, and width) to construct a viewing geometry that
is mapped to changes in plasma conditions over the \1624 orbital
period.

\end{abstract}


\section{Introduction} 
The ``Big Dipper'' \1624, with its high energy ($>8$\,keV) flaring
between dips, is one of the most `extreme' of the dipping sources. Its
nickname derives in part from the fact that it has one of the longest
orbital periods (21\,hr $\approx 76$\,ks; \citealt{jones89}) and dips
that last $\approx 2.7\,\rm hrs \approx 11$\,ks \citep{watson85,
  balucinska00}.  The prevailing idea for explaining this behavior is
that the dips are likely due to obscuration by a raised accretion disk
rim at the location where the infalling accretion stream impacts the
disk. By accounting for the quasi-steady X-ray dust
  scattering halo, one can show that obscuration during dips is
  $>$90\% \citep{xiang07, iaria07}. The dips exhibit broad, shallow
shoulders, with a rapid ingress/egress to the deepest dip levels. A
point-like soft X-ray source reasonably can be inferred from the rapid
ingress/egress and dip variability. While the dipping seems to
saturate at a minimum level, rapid ($<30$~s), short duration, large
amplitude spikes are seen. This behavior argues for the obscuring
medium having a well-defined boundary, but a highly variable, erratic
covering.

Since its discovery \citep{watson85}, \1624 has been studied
extensively with X-ray telescopes including
\exosat\ \citep{church95}, \ginga\ \citep{jones89},
\einstein\ \citep{christian97}, \asca\ \citep{angelini97, asai00},
\beppo\ \citep{balucinska00}, \xte\ \citep{balucinska01, lommen05,
  smale01}, \xmm\ \citep{parmar02, trigo06}, and
\chandra\ \citep{iaria07, xiang07, wachter05}. From these studies we
know that \1624 is highly absorbed, both extrinsically \emph{and}
intrinsically.  Models have been developed over the years wherein the
spectrum consists of: large external absorption, a scattering halo, a
point-like (i.e., rapidly absorbed) blackbody, and an extended (i.e.,
partially absorbed) Accretion Disk Corona (ADC) \citep{balucinska00,
  balucinska01, smale01}. Observations also have revealed a
\emph{broad} Fe K$\alpha$ emission line \citep{smale01, parmar02}. It
is therefore likely that we are viewing both the inner (point-like
source, broad lines) and outer (dips, variable absorption) accretion
disk regions in this highly inclined, high Eddington fraction,
low-mass X-ray binary (LMXB).

From the point of view of high resolution \chandra\ spectroscopy,
\1624 may be in many ways analogous to the bright, highly inclined,
variable, and absorbed microquasar \grs1915
\citep{lee02,grs1915nature}. As such, there are key questions
concerning the physical environment of \1624 that \emph{only} high
resolution X-ray spectroscopy can address. Is the extended corona a
photoionized atmosphere in equilibrium? Medium resolution \xmm\ EPIC
data provide some clues. \citet{parmar02} identified `narrow'
($<$50\,eV) resonant absorption lines from Fe~{\sc xxvi}~Ly$\alpha$,
Fe~{\sc xxvi}~He~$\alpha$, and a broad(er) 470~eV Fe~K$\alpha$
emission line (see also \citealt{smale01}). Like all X-ray sensitive
CCD cameras, however, the \xmm\ EPIC, with a resolution comparable to
the \asca\ SIS, is fundamentally limited in the extent to which it can
resolve such features.  A recent analysis of these \chandra\ grating
observations by \cite{iaria07}, while providing a better measurement
of the ionized Fe lines given the better HEG resolution compared with
EPIC in this spectral band, did not undertake the in-depth modeling
necessary to describe the viewing geometry for the photoionized
gas. Furthermore, based on the analysis presented in this paper, we
offer an alternative scenario to the interpretation of Iaria et
al.\ that the observed absorption line width can be attributed to bulk
Comptonization.

As a prelude to this paper, we studied the dust scattering halo
properties of \1624 with this \chandra\ High Energy Transmission
Grating Spectrometer (HETGS) observation in an attempt to first better
understand the distribution of intervening line-of-sight material
(size and spatial distribution of dust grains), which we found to be
heterogeneously distributed near the source and in the spiral arms of
the Galaxy \citep{xiang07}. Additionally, by measuring scattering
delay times as a function of angular distance in comparison to the
\1624 source light curve, we were able to obtain a much more accurate
distance measurement of $15_{-2.6}^{+2.9}$~kpc to \1624. In this
paper, we use the same \chandra-HETGS observation to focus on the
binary properties with an in-depth analysis of the HETGS spectra.
(These are the same data published by \citealt{iaria07}.)
Specifically, we explore the persistent-phase spectrum over the \1624
orbital period in the context of \S\ref{sec-continuum}:
photoionization modeling using {\sc xstar} \citep[][Version
  2.1kn7]{xstar}, and \S\ref{sec-absorption}: detailed variability
studies of the detected ionized \ion{Fe}{25} and \ion{Fe}{26}
absorption lines and continuum. In particular, we study the orbital
phase evolution of the detected absorption lines as a function of the
time periods immediately before and after (\S\ref{sec-prepost}), as
well as near to and far from the dipping events (\S\ref{sec-nfdip}),
in addition to an analysis of the short-duration ($\approx 10$~ks)
absorption changes over the \1624 $\approx 21~\rm hr$ orbital period.
The results are summarized and discussed in \S\ref{sec-discussion} in
the context of a simple two temperature absorber model, and we
specifically focus on the quasi-sinusoidal modulation of these plasma
components, as well as modulation of the continuum components, over
the binary orbital period.

\section {Observations and Data Analysis}
\label{sec-observation} We observed \1624 with the \chandra\ High
Energy Transmission Grating Spectrometer (HETGS,
\citealt{canizares05}) in timed graded mode beginning at 06:26:06 UT on 
2004 June 04 (MJD 53160.26813, ObsID: 4559), covering one binary orbit of
$\approx$76~ks. Given that \1624 is bright ($\approx$50\,mCrab,
\citealt{smale01}, \citealt{trigo06}), we mitigate pileup and
telemetry problems by using a one-half subarray of 512 columns per CCD
in order to reduce the frame-time to 1.7~s from the nominal 3.2~s.
Despite these efforts, the 0$^{th}$ order ACIS-S data are heavily
piled-up, rendering difficult the drawing of useful conclusions from
the 0th order spectrum. Accordingly, this paper focuses only on spectral and
light curve analysis of the HETGS grating spectra.

We use {\sc ciao}~3.4 with {\sc caldb}~3.3 for our data reduction
efforts. In order to ensure the most accurate wavelength measurements
possible for this observation, we calculate the exact zero order
position using a code that fits the intersection of the dispersed MEG
data with the zeroth-order ``readout trace'' on the ACIS-S3
chip\footnote{http://space.mit.edu/cxc/analysis/findzo/}. Using this
technique it is expected the accuracy of the 0$^{th}$ order position
is better than 0.1 detector pixels, translating to a wavelength
accuracy of 0.001\,\AA\ (corresponding to 164~km~s$^{-1}$ at 6.8~keV) for both
the High Energy Grating (HEG) and Medium Energy Grating (MEG) first
orders.  Positive and negative first order spectra for HEG and MEG
were combined for analysis.

\begin{figure}
\plotone{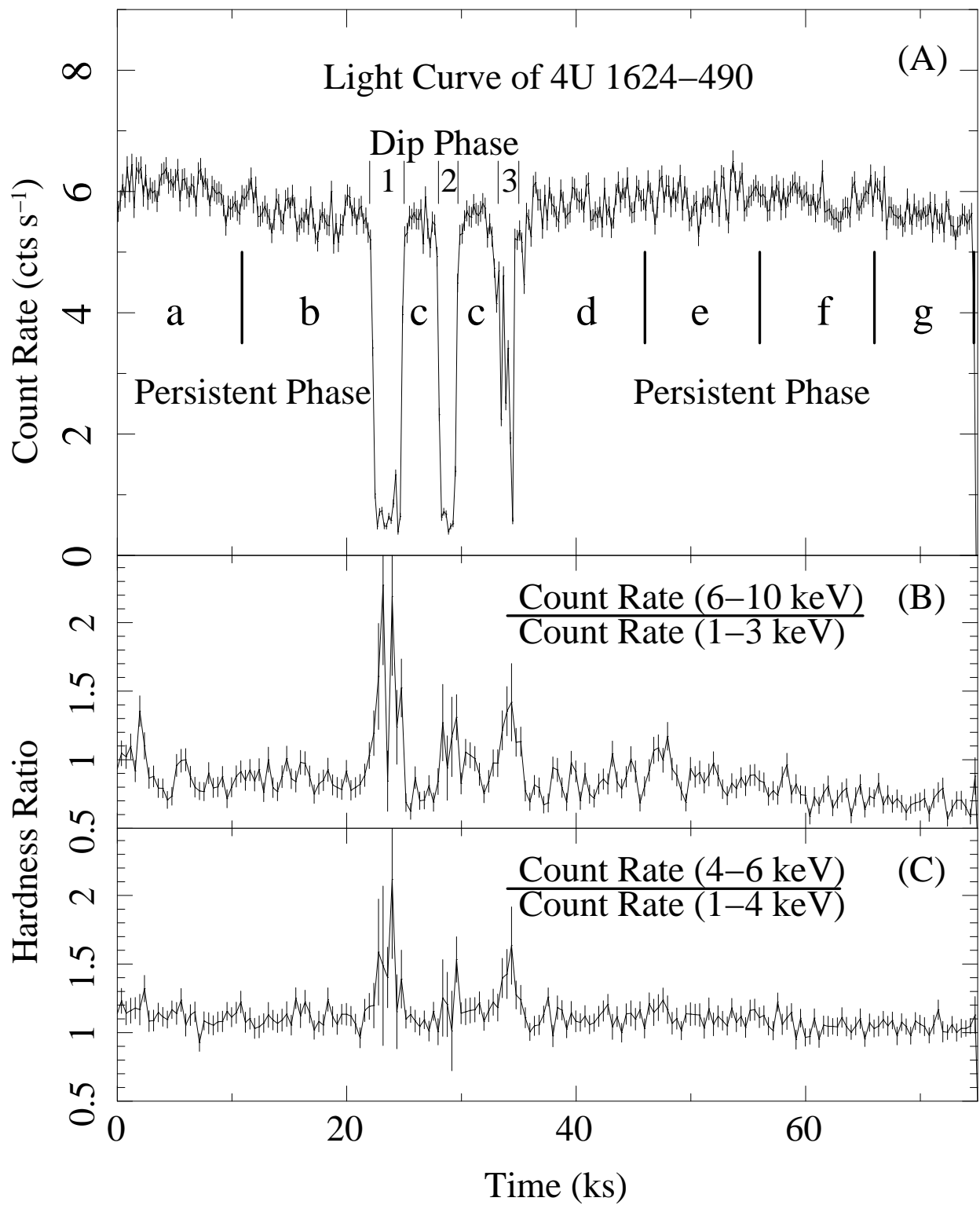}
\caption{(A) First-order 1--10~keV HEG+MEG light curve of
  4U~1624$-$490, and (B,C) associated hardness ratios that reveal 
  spectral hardening during the dip periods. Throughout the paper, 
time zero corresponds to MJD 53160.3.}
\label{fig-lc}
\end{figure}

For our analysis, we generate $1^{\rm st}$-order time-averaged HEG and
MEG spectra spanning the non-dip phases of the full 76~ks orbital
period (resulting in $\approx$67~ks of usable data) to study the
properties of the persistent phase continuum spectrum and to look for
faint lines (\S\ref{sec-continuum}).  We also generate phase-dependent
spectra (labeled ``a--g'', and combinations thereof, as indicated by
the light curve in Fig.~\ref{fig-lc}) to study the time evolution of
the plasma via observed changes in the line profiles and fluxes
(\S\ref{sec-absorption}). Figure~\ref{fig-lc} also shows that our
observation encompasses $\approx$9.7~ks ($\approx$2.7 hrs; sum of dip
phases 1+2+3) worth of occulted periods. The longest duration of an
individual dip is seen to be $\approx$3.5~ks, and the largest
occultation is $\approx$90\%.  Possibly because of the low energy
absorption, the spectrum is seen to harden during dipping periods. We
are unable to extract useful information from the {\it dip-phase}
HETGS spectrum due to the low counts.

\subsection{The Average Persistent Phase Spectrum}
\label{sec-continuum} As a prelude to spectral analysis, we first
determined pileup fraction (defined to be the fraction of frames that
have detected events containing two or more events per frame per
pixel) in the first order grating spectra. Since pileup is dependent
on the count rate in each CCD pixel, we use the count rate in the
HEG$\pm$1 and MEG$\pm$1 spectra for our estimates. We find that the
maximal pileup fraction of HEG +1 and $-1$ order spectra is
$\approx$1.0\% at $\approx$2.5~\AA\, ($\approx$5~keV). Similarly MEG
+1 and $-1$ order spectra are piled up at less than $\approx$1.0\%
except for the range 2.5--3.5~\AA\, (3.5--5.0~keV) where the maximal
pileup fraction reaches $\approx$2.0\%. Accordingly, we ignore regions
of the spectrum with pileup $>$1\%. As such, we fit only to 1.2--7.3~\AA\,
(1.7--10.0~keV) HEG spectra, and MEG spectra over the 1.5--2.5~\AA\,
(1.7--3.5~keV) and 3.5--7.5~\AA\, (5.0--8.0~keV) range. The spectra
are rebinned to satisfy a signal-to-noise ratio $S/N>5$ and 0.005 \AA\
per bin for the HEG 1$^{\rm st}$ order, $S/N>5$ and 0.01~\AA\ per bin 
for the MEG 1$^{\rm st}$ order, respectively. {\sc
  ISIS}\footnote{http://space.mit.edu/cxc/isis/} \citep{houck00,
  noble08} was used as the fitting engine for our analysis.

\begin{figure}
\epsscale{0.9}
\plotone{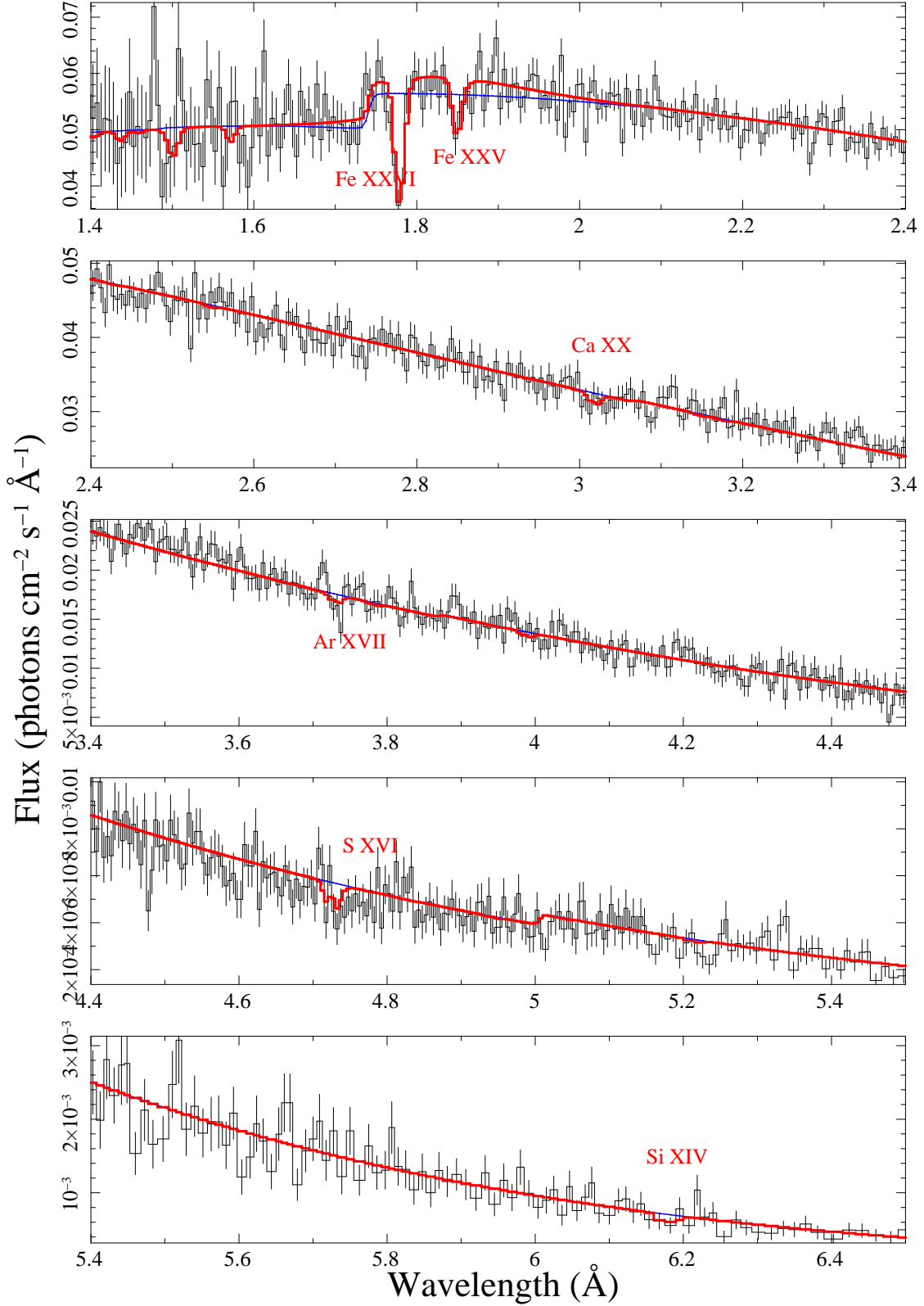}
\caption{The best-fit persistent phase continuum and broad Fe
  emission line model (red) of Table~\ref{tab-cont} modified by two
  photoionized absorption lines (see \S\ref{sec-line}) overplotted on
  the time-averaged HEG $1^{\rm st}$ order spectrum of \1624 over its
  $\approx 76$~ks orbital period. The blue line is the best-fit model
  excluding the Fe emission line and the ionized absorption lines. }
\label{fig-continuum}
\end{figure}

\subsubsection{Broadband Continuum}
\label{sec-broad}

The \1624 broadband continuum as determined from the HEG spectrum has
already been discussed at length in our previous paper focusing on the
scattering halo associated with this source \citep{xiang07}. The
inclusion of the MEG spectrum to our fitting here gives consistent
results for a blackbody + power-law continuum modified by absorption
from neutral material ($N_{\rm H}$). These parameters are listed in
column 3 of Table \ref{tab-cont}, and a spectrum is shown in
Fig.~\ref{fig-continuum}. In our previous work, however, we were
concerned by the fact that our best-fit parameters were noticeably
different from that published by \citet{iaria07} for the same data
set. In particular, our best-fit photon index $\Gamma
=1.14_{-0.48}^{+0.33}$ is slightly harder than the
$\Gamma=2.0_{-0.55}^{+0.30}$ noted by Iaria et al. A cursory check
showed that the difference had very little to do with absorption
models, e.g., $\Gamma \approx 1.4$ for {\sc phabs}
\citep{balucinska92}, versus $\Gamma \approx 1.2$ for {\sc tbabs},
which use the updated cross-sections and abundances of
\citet{wilms00}, fitting tools (e.g., {\sc isis} versus {\sc xspec}),
binning, or choice of fitted spectral regions. For completeness, we
also looked into calibration effects and find that at least some of
the differences had to do with a difference between {\sc caldb 3.2}
(used by Iaria et al.) versus {\sc caldb 3.3} (this paper). A
reduction of the \1624 spectra using {\sc caldb 3.2} resulted in a
slightly softer power-law ($\Gamma \approx 1.8$), compared to our {\sc
  caldb 3.3} value of $\Gamma \approx 1.4$ ({\sc phabs}). Even though
the {\sc caldb 3.2} measure of photon index is closer to the
$\Gamma=2.0_{-0.55}^{+0.30}$ value quoted by Iaria et al.\ for this
data set, a full agreement with the values of Iaria et al.\ is not
reached even when trying to reproduce every step of their analysis
process. We can only speculate that a combination of effects,
including new reprocessing that accounts for afterglow events, and the
careful treatment of pileup incorporated here may be responsible for
the rest of the discrepancy. We also note that our
  photon index is slightly harder than the values of
  $\Gamma=2.0_{-0.8}^{+0.5}$ for a \beppo observation by
  \citet{balucinska00} and $\Gamma=2.02\pm0.12$ for a \xmm observation
  by \citet{parmar02}. The difference is possibly caused by the
  different luminosity states (see the next paragraph) or by the
  limited \chandra-HETGS band coverage (e.g. 0.5--10~keV) and the
  severe absorption at the low energies coupled with limited counts at
  the high energies.

Based upon our best-fit continuum, the 0.6--10\,keV flux of \1624 is
$\approx1.07\times 10^{-9}$~erg~cm$^{-2}$~s$^{-1}$
($\approx50$~mCrab), corresponding to an unabsorbed 0.6--10 keV
luminosity of $\approx4.9\times10^{37}$~ergs~s$^{-1}$ (assuming a
distance of 15.0~kpc as determined from our scattering halo analysis;
\citealt{xiang07}). The luminosity is similar to that observed with
\xmm\ \citep{parmar02} but significantly lower than that using the
\xte\, \citep{smale01} and \exosat\ \citep{church95} observations.
       As discussed by \citet{smale01}, the luminosity
         during different observations can change significantly and
         thus the emission parameters should not necessarily be
         expected to completely consistent.

As an alternative to the continuum model of a blackbody plus power-law,
we also considered a continuum consisting solely of a partially
covered power-law. In this model, the soft excess that had been
described by the blackbody is now modeled by the soft end of the
uncovered power-law. Specifically, the model is given by
\begin{equation}
f_{\rm obs}(E) = A e^{-N_{\rm H,\ LOS}\sigma_{\rm ph}(E)}
\left(1-x + xe^{-N_{\rm H,\ Local}\sigma_{\rm ph}(E)}\right) 
E^{-\Gamma}
\end{equation}
where $N_{\rm H,\ LOS}$ is the equivalent hydrogen column along the
line-of-sight, $N_{\rm H,\ Local}$ is the equivalent hydrogen column
local to the system, $A$ is the normalization of the power-law
continuum, and $x$ is the partial covering fraction. We find that this
model ($\chi^2/d.o.f=1413/1338$) describes the persistent phase
spectrum slightly better than the blackbody plus power-law model
($\chi^2/d.o.f=1435/1338$), and yields values of
  $x=(71\pm3)\%$ , $\Gamma=2.25\pm0.06$ and $N_{\rm
    H,\ Local}=8.1_{-0.6}^{+0.7}\times10^{22}\, \rm cm^{-2}$.  Again,
  as discussed above, the limited \chandra-HETGS band coverage
  (e.g. 0.5--10~keV) and the limited counts at high energies do not
  allow us to distinguish between these two models for the time
  averaged spectra. Thus for some of the results that follow, e.g.,
  the modeling of the broad emission line (\S\ref{sec-bline}) and
  narrow ionized absorption lines (\S\ref{sec-line}), the results are
  insensitive to which continuum model that we choose.  As we
  elaborate upon in the discussion of short time scale changes
  (\S\ref{sec-sdchange}), however, the partial covering model may more
  naturally explain spectral variations associated with the source's
  lightcurve.

\input{t1.tab}

\subsubsection{Broad Emission Line}\label{sec-bline} 
A significant and broad excess around 6.6~keV was found when we
checked the fit residuals in both the HEG and MEG spectra. We also saw
deficits around 6.7~keV and 7.0~keV corresponding to narrow absorption
lines. These are discussed in more detail in
\S\ref{sec-line}~and~\S\ref{sec-absorption}. The broad excess has been
identified with Fe K emission typically detected in LMXBs ($1^{\rm
  st}$ significant detection: \citealt{suzuki84} with {\it Tenma};
with {\it EXOSAT}, \citealt{white86} and \citealt{hirano87};
\citealt{asai00} for an {\it ASCA} study of 20~LMXBs, including
\1624). \citet{asai00} postulate that the line is likely produced
through the radiative recombination in a photoionized plasma and that
its broad width is attributable to a confluence of processes in the
ADC, including Compton scattering and Doppler shifts arising from
Keplerian motions. \citet{parmar02} also studied the broad emission
line from \1624 with \xmm, and found that the best-fit energy for the
emission line depends on whether or not the Fe absorption lines were
also modeled. When the absorption lines were included,
\citeauthor{parmar02} found a best-fit emission energy of
$6.58_{-0.04}^{+0.07}$~keV in contrast to the
$6.39_{-0.04}^{+0.03}$~keV value found when the absorption features
were excluded. The HEG clearly shows strong \ion{Fe}{25} and
\ion{Fe}{26} absorption lines in our HEG spectrum, which are therefore
included in our modeling of the the broad emission feature. The
continuum model is the blackbody plus power-law model of
\S\ref{sec-broad}.

Like the persistent phase spectrum, the far-dip ({\it a+e+f+g};
$\approx$39~ks) and near-dip ({\it b+c+d}; $\approx$29~ks) spectra of
Fig.~\ref{fig-line2} reveal a broad emission line with a best-fit
energy that is most consistent with He-like \ion{Fe}{25} at $\approx
6.7$~keV (Table~\ref{tab-cont}). It is interesting to note that the
peak energy for the broad emission feature in the near-dip phase
spectrum points to even higher ionization H-like \ion{Fe}{26} and to a
broader velocity width, although the uncertainties of the modeling do
not allow us to make any strong claims to this effect. If real, it
would be of interest to consider whether Compton scattering can be a
mechanism for line broadening in LMXBs as suggested by
\cite{asai00}. Note that the highest values for the
  fitted line width, $\approx 0.5$\,keV, correspond to velocities that
  if associated with the accretion disk arise at radii $>100~GM/c^2$.
  Thus, the broad emission line is not necessarily cospatial with the
  blackbody component, if indeed a blackbody+powerlaw is the better
  description of the true, underlying physical description of the
  observed spectra.  The presence of a broad line therefore does not
  provide us with a means of distinguishing between the
  blackbody+powerlaw and partially covered powerlaw models for the
  continuum spectra, and could be consistent with either model.

\subsubsection{Photoionization modeling of the narrow ionized absorption lines}
\label{sec-line} The three most obvious absorption lines that show up
in the \chandra-HETGS spectra (\ion{Fe}{26} at
$\approx$~7.0~keV/1.778~\AA\, \ion{Fe}{25} at
$\approx$6.7~keV/1.850~\AA\, and \ion{Ca}{20} at
$\approx$4.1~keV/3.020~\AA; see Fig.~\ref{fig-continuum}) have already
been reported by Iaria {\it et al}, albeit based upon
  Gaussian fits to these data, as opposed to the modeling with
  photoionization codes as we present in this work.  Ionized
absorption lines were initially discovered for \1624 in \xmm EPIC data
\citep{parmar02}. Here, for the time-averaged spectra we go beyond
previous studies by modeling the absorption spectrum of \1624 using an
analytic version of {\sc xstar}~v2.1ln2 \citep{xstar}. For these {\sc
  xstar} fits, the column density, ionization, turbulent velocity width,
and velocity shift were allowed to vary. The abundance of elements
from C to Fe were fixed at solar values \citep{grevesse96}, although
subsequent fits that allowed these parameters to vary gave abundance
results similar to solar values within 90\% confidence errors. If we
adopt the interpretation of Iaria et al.\ and consider a one-zone
plasma, the time-averaged spectrum over the course of the \1624
orbital period can be modeled reasonably well with a hot plasma of
$\log\,\xi = 3.6 \pm 0.2$ ionizing a column of log~$N_{\rm H}=
22.53_{-0.08}^{+0.07}\,{\rm cm}^{-2}$, with turbulent velocities of
$v_{\rm turb} = 656_{-334}^{+480}\, \rm km \, s^{-1}$. Because our
variability analysis of \S\ref{sec-absorption} strongly points to a
two-ionization zone model for the \1624 absorbers, we describe our
model for the time-averaged spectrum accordingly. For this model, a
combination of ionized absorber-1 ($\rm log\,\xi =4.3\pm 0.4$; $\log
N_{\rm H}= 23.3\pm \rm 0.2$; $v_{\rm turb} = 280_{-80}^{+180}\rm km \,
s^{-1}$, and $v_{\rm shift}=-607_{-342}^{+354}\rm km~s^{-1}$), and
absorber-2 ($\rm log\,\xi =3.3\pm0.2$; $\log N_{\rm H}=
22.1_{-0.2}^{+0.1}\rm cm^{-2}$; $v_{\rm turb} < 174\, \rm km \,
s^{-1}$ and $v_{\rm shift}=213_{-158}^{+108}\, \rm km \, s^{-1} $),
provides a better fit. Interestingly, for either scenario the
\ion{Ar}{18} component cannot be modeled well unless the abundance of
Ar is set to $\approx$twice solar. 
  Table~\ref{tab:xstar} lists ions detected in our spectrum, and also
  the strong lines that are predicted by our {\sc xstar} models to have
  equivalent widths $W_{\rm ion} > 0.1$~m\AA, including three strong
  lines of \ion{Fe}{26}, \ion{Fe}{25} and \ion{Ca}{20}.

\input{t2.tab}

\subsection{Orbital Phase Evolution of the H-like and He-like Fe Absorption}
\label{sec-absorption} Given that the light curve (Fig.~\ref{fig-lc})
clearly indicates a blockage of the primary by the accretion stream
and/or secondary, we next investigate relevant effects this might have
on the observed ionized absorption. For this, we concentrate only on the
evolution of the strongest absorption lines, i.e., \ion{Fe}{25} and 
\ion{Fe}{26}, over the course of the \1624 $\approx$~76~ks orbital
period, as the source moves in and out of occultation. For this
analysis, we use only the HEG given its higher resolution and
throughput in this energy band.

\subsubsection{Pre-dip versus Post-dip phases Fe line changes}
\label{sec-prepost}
A simple overlay of the pre-dip (\textit{a+b}; $\approx$20~ks) and
post-dip (\textit{d+e+f+g}; $\approx$38~ks) phases reveal nearly
identical spectra. Our analysis here uses the blackbody plus power-law
continuum (\S\ref{sec-continuum}) and broad Fe emission line 
(\S\ref{sec-bline})  to model
the more global features.  While the underlying components are similar
to that used for the time averaged spectrum, the best-fit parameters
differ slightly. To quantify this effect, we assess Fe line changes
by adding Gaussians to model the \ion{Fe}{25} and \ion{Fe}{26}
absorption. The best-fit parameters (Table~\ref{tab-gauss}) do not
indicate significant changes in absorption between the {\it pre-dip}
and {\it post-dip} phases, so for these fits we do not proceed with
the more involved {\sc xstar} modeling.

\subsubsection{Near-dip versus far-dip phases Fe line
  changes} \label{sec-nfdip} 

\begin{figure}
%\epsscale{1.0}
\plotone{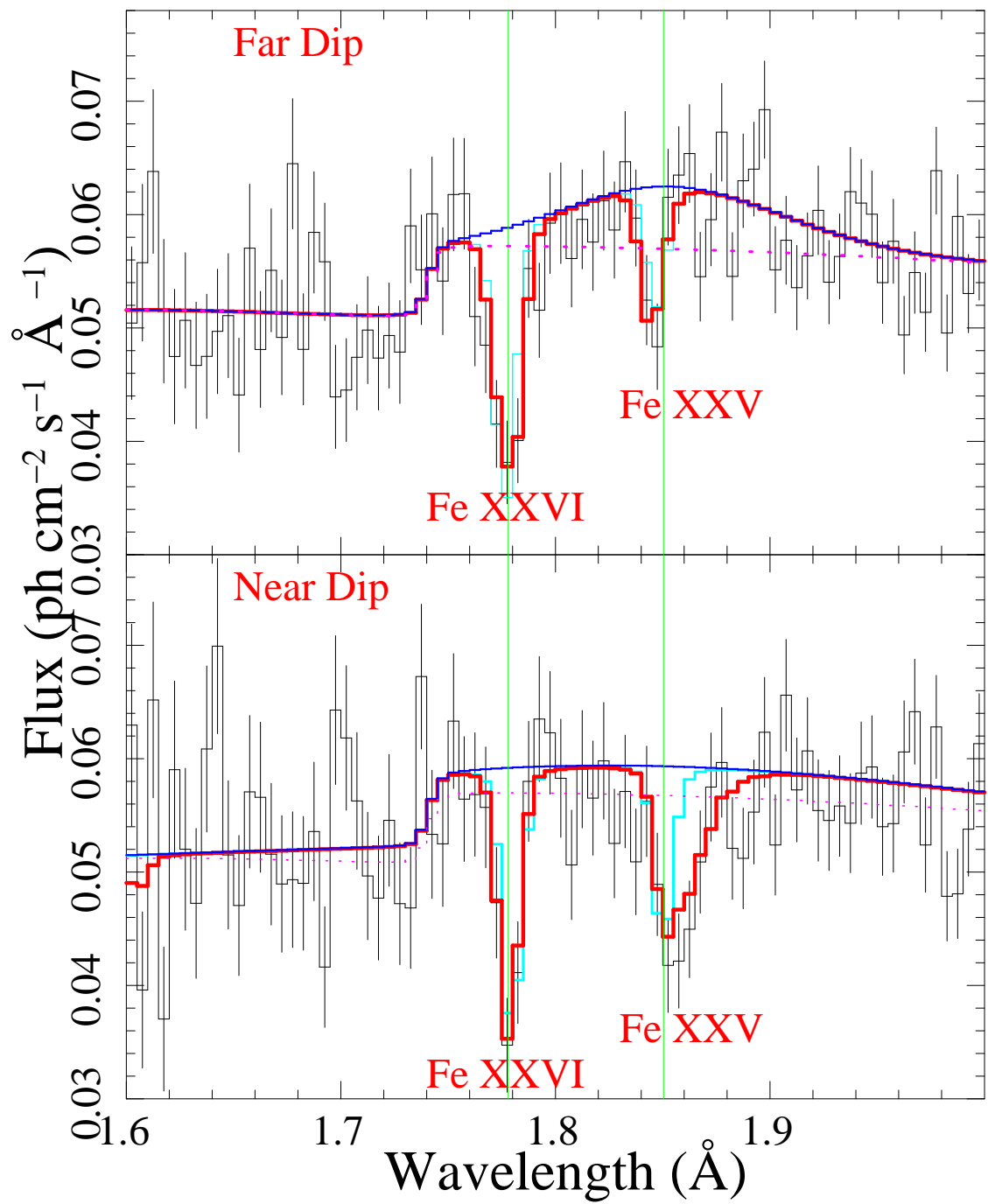}
\caption{Fluxed spectra (black) and best-fit two-component ionization
  model (red) of \S\ref{sec-nfdip} during the far-dip (top) and
  near-dip (bottom) \1624 phases. Represented in light blue is a
  single-component ionization model that clearly fits the data less
  well, in dark blue is the model excluding the ionized
  absorption lines. Vertical green lines indicate the laboratory
  wavelengths of \feII (1.850\AA) and \feI (1.778\AA). }
\label{fig-line2}
\end{figure}

Next, we considered a comparison between the {\it near-dip} ({\it
  b+c+d;} $\approx29$~ks) and {\it far-dip: (a+e+f+g}; $\approx
39$~ks) spectra following (initially) the procedure of
\S\ref{sec-prepost}. Based on simple Gaussian fits, distinct Fe line
changes are observed in the line parameters, most noticeably in the
\ion{Fe}{25} velocity width, equivalent width, and line flux
(Fig.~\ref{fig-line2} and Table~\ref{tab-gauss}). Fig.~\ref{fig-line2}
and Table~\ref{tab-gauss} reveal distinct Fe absorption line changes
with the \ion{Fe}{25}:\ion{Fe}{26} flux ratios indicating noticeable
plasma changes (temperature \& ionization) between the {\it near-dip}
and {\it far-dip} phases.

In an attempt towards a more physically realistic picture explaining
these changes, we proceed with more rigorous photoionization modeling
of these spectra. For this, we fixed the continuum parameters specific
to each phase, and varied the {\sc xstar} parameters as before to
model the \ion{Fe}{25} and \ion{Fe}{26} absorption features. For these
spectra, we find that a single ionization absorber gives an overall
reasonable fit in a $\chi_\nu^2$ sense, but does not fully describe
the Fe {\sc xxv} absorption line. The fact that the relative
\ion{Fe}{25} and \ion{Fe}{26} velocity shifts are different in the
far-dip and near-dip spectra (i.e., Fig.~\ref{fig-line2}) suggests
that a more realistic physical model should consider two ionized
absorbers arising from different physical regions.  A fit to these
near-dip and far-dip spectra using such a two-component ionized
plasma: one hot ($\approx3.0\times 10^6$~K corresponding to $\log \xi
\approx 4.2$) and another slightly cooler ($\approx1.0\times 10^6$~K;
$\log \xi \approx 3.0$), gives a better description
(Fig.~\ref{fig-line2}). By adding this second
  component, the $\chi^2$ of the fit to the near-dip phases is
  reduced from 1047 for 1063 d.o.f (one-component ionized plasma) to
  1033 for 1059 d.o.f (two-component ionized plasma).  This yields an
  F-statistic value of 3.6, which for these four additional degrees of
  freedom formally corresponds to an improved fit at $>99\%$
  confidence.  (See, however, \citealt{protassov:02a} for a discussion
  of the applicability of the F-test in situations such as those
  presented here.)  We therefore view the two-component ionized
plasma model as a more likely scenario for the \1624 X-ray emitting
regions.

\input{t3.tab}
\input{t4.tab}

\subsubsection{Short duration changes} \label{sec-sdchange} 
\begin{figure}
%\epsscale{2.0}
\plotone{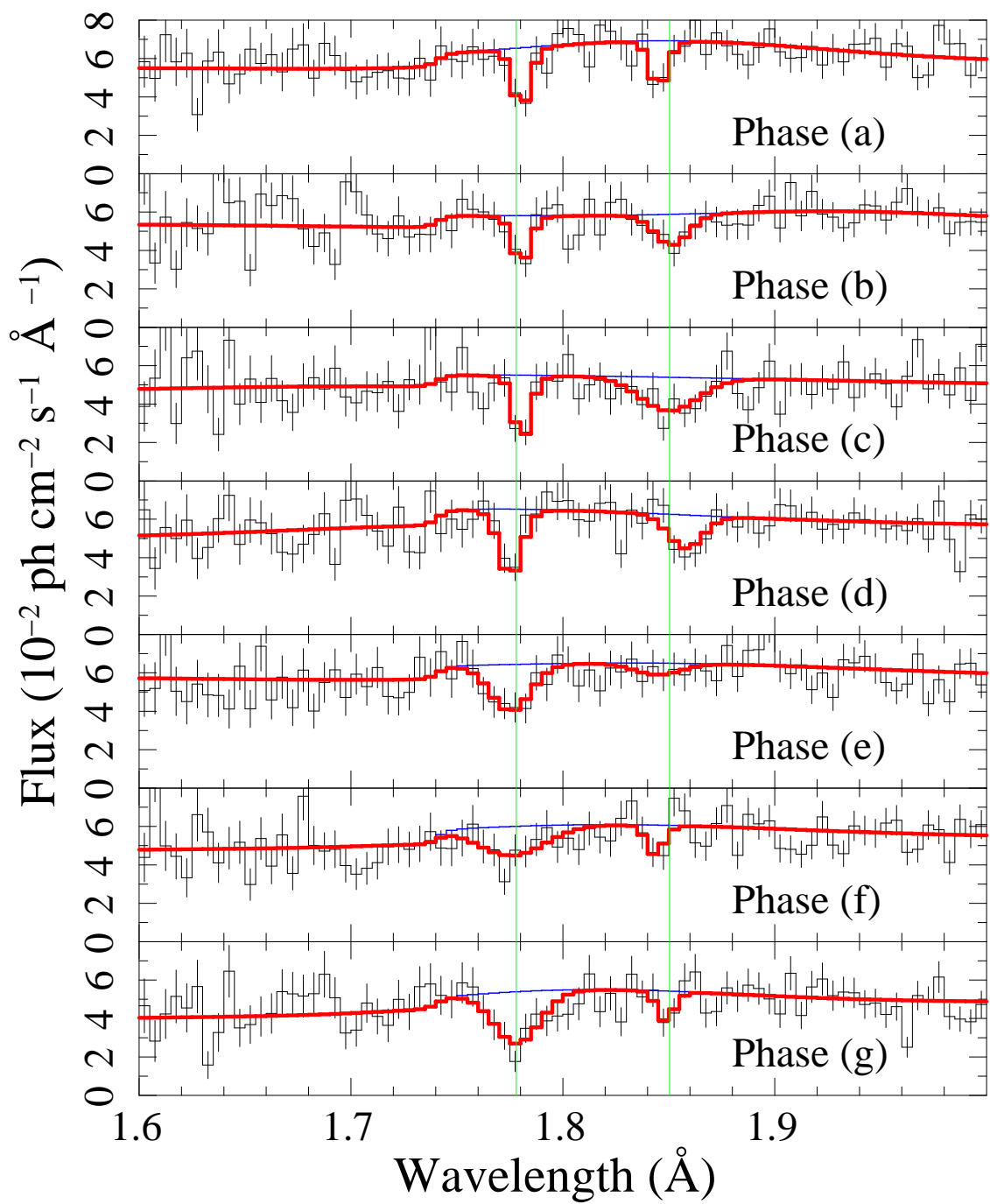}
\caption{HEG spectra (black) and best-fit model (red) showing the
  evolution of the \ion{Fe}{25} and \ion{Fe}{26} absorption lines over
  the \1624 orbital period. As in Fig.~\ref{fig-line2}, the vertical green
  lines indicate the laboratory wavelengths of \feII (1.850\AA) and
  \feI (1.778\AA).}
\label{fig-line}
\end{figure}

\begin{figure}
\epsscale{0.8}
\plotone{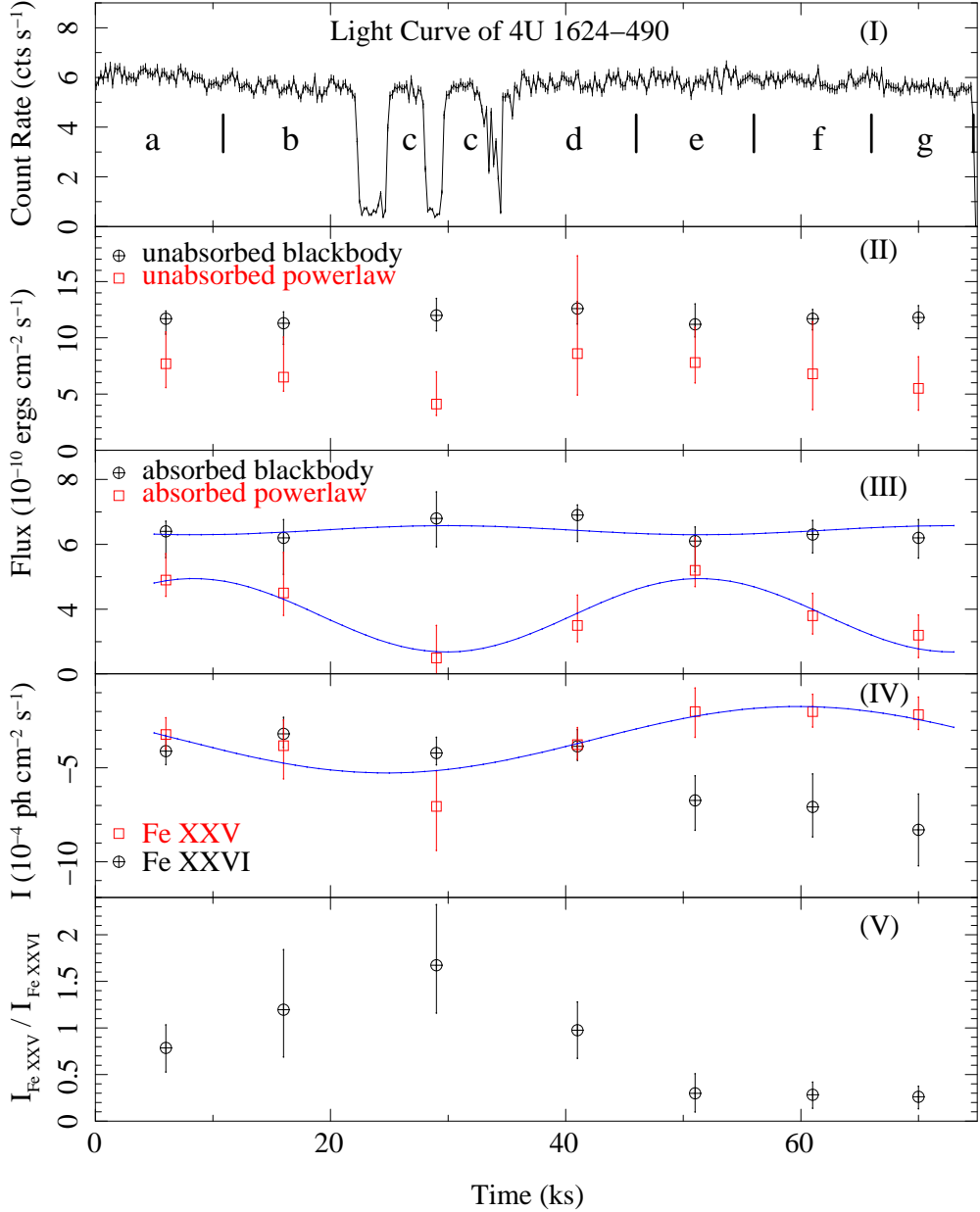}
\caption{(I) Tracking the \1624 light curve over 10~ks time steps,
  shows that the unabsorbed fluxes of both the blackbody and
  power-law components (II), and to a lesser extent the absorbed flux
  of the blackbody component (III), remain relatively steady, whereas
  the absorbed power-law component (III) reaches a minimum during the
  periods near dipping events. Both components can be described by
  sine waves (blue curves). (IV) Intensity evolution of the Fe absorption 
  lines. The blue curve is also a sine wave fitting to the intensity 
  of \ion{Fe}{25} absorption line. (V) The evolution of the
  intensity ratio of the absorption lines shows a quasi-sinusoidal
  variation. In this figure only, all error bars shown are at the
  $1\sigma$ confidence level. The vertical bars in subfigure (I)
  indicate the different phases of flux evolution.}\label{fig-evo}
\end{figure}

Since
there is clear evidence for absorber changes when comparing the far-
and near-dip periods, we conclude our analysis of the spectral
variability by dividing the total persistent spectra into 7
approximately equal duration parts of $\approx$10~ks each
(Figs.~\ref{fig-lc}, \ref{fig-line}~and~\ref{fig-evo}).  Of interest
is whether the blackbody+power-law or partial
  coverer+power-law components vary with orbital phase. We first test
for such variations by fitting our blackbody plus power-law
continuum model, including the broad emission line and narrow Fe
absorption lines. Fig.~\ref{fig-evo}-(II, III) and
Table~\ref{tab-7con} show that the unabsorbed fluxes of both the blackbody 
and power-law components, and to some extent the absorbed flux of
the blackbody component, remain relatively steady over the \1624
orbital period.  The absorbed power-law component, however, is seen to
vary more dramatically, and in a quasi-sinusoidal fashion ($\approx$95\%
confidence).

We note that the quasi-sinusoidal variation of the absorbed blackbody
component is less than that of the absorbed power-law, and perhaps
exhibits an anti-correlation with the power-law.  We therefore assess
the statistical correlation between these parameters by generating
confidence contours for phase (g) (see Fig \ref{fig-line} and \ref{fig-evo}).
%where Fig.~\ref{fig-evo}-(III) shows these two flux components to have the
%greatest anti-correlation.  
These contours indicate a marked degree of
dependence of the absorbed power-law and blackbody fluxes upon each
other; therefore, in deriving a period for the modulation of the flux,
we tie these two components together and fit a sine function to the
points of Fig.~\ref{fig-evo}-(III). Specifically, the power-law and
blackbody fluxes are tied together, but the power-law flux is allowed
a 1/2 period (delayed) phase shift. At 90\% confidence, we derive a
period $T=43_{-9}^{+13}$~ks ($\approx$~12~hrs), slightly more than half the
orbital period.

\input{t5.tab}

\begin{figure}
%\epsscale{0.7}
\plotone{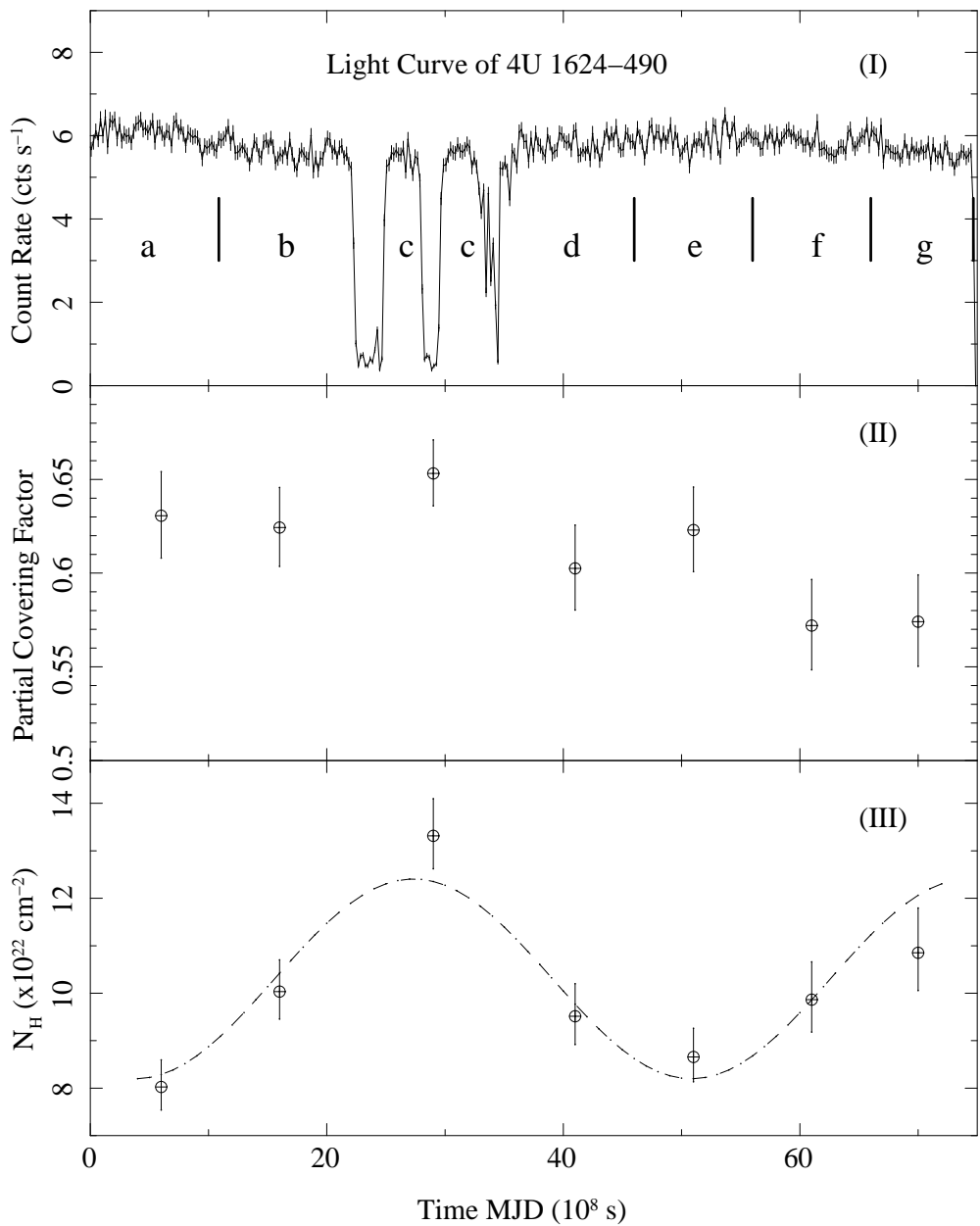}
\caption{The evolution of (II) the partial covering factor and 
(III) the equivalent hydrogen column density of the local absorber, which
reaches a maximum during the periods near dipping events. Similar to the 
fits in Fig \ref{fig-evo}, the equivalent hydrogen column density also 
can be described by a sine wave with a period of $T=47\pm7$~ks (dashed curve).
In this figure, all error bars shown are at the $1\sigma$ confidence level. }
\label{fig-partial}
\end{figure}

 Next, we explore the time dependence of the partial
  covering model (see the description in \S\ref{sec-broad}) by fitting
  simultaneously the seven phased spectra.  We wish to determine
  whether the change of the absorbed power-law flux is produced by
  variation of the local cold gas. Here we tie together all fit
parameters (including power-law normalization and slope), but allow
the partial covering fraction and local absorption, i.e, the cold
absorption along the line-of-sight, to vary with orbital phase.
We find that this model describes the seven phased spectra well,
giving $\chi^2$ of 5971 for 5896 degrees of freedom and yields values
of $N_{\rm H}=9.11_{-0.24}^{+0.23}\times10^{22}\ \rm cm^{-2}$ and
$\Gamma=2.26_{-0.05}^{+0.06}$.  Figure~\ref{fig-partial} shows that
the equivalent hydrogen column density of the local cold absorber
varies significantly over the orbital phase, while the partial
covering factor remains relatively steady. This variation of the
hydrogen column density of the local cold absorber is very similar to
the phase-dependent variations seen for the absorbed power-law flux in
the alternative blackbody+power-law models described
above; therefore, we also use a sine function to fit the points of
Fig.  \ref{fig-partial}-(III). At 90\% confidence, we derive a period
$T=47\pm7$~ks, similar to the fits above. Again, we are seeing evidence
that the observed orbital modulations are predominantly driven by
absorption variations, rather than by intrinsic source variations,
regardless of whether we fit a blackbody plus power-law model, or
instead fit a partial covering model of a single component (i.e., the
power-law).

We also investigate \ion{Fe}{25} and \ion{Fe}{26} absorption line
changes over these smaller time segments by fitting Gaussian lines to
the absorption features while fixing the continuum parameters to the
phase-specific values of Table~\ref{tab-7con}.  (Here, the HEG
spectrum is binned to 0.005~\AA\ bin$^{-1}$.)  Table~\ref{tab-Fe} and
Figs.~\ref{fig-line}~and~\ref{fig-evo}-(IV) show that the \ion{Fe}{26}
velocity width is narrowest during phases~{\it c}~and~{\it d} that
occur immediately after dipping events. In contrast, it appears that
it is the \ion{Fe}{25} flux and velocity shift that change most
dramatically over the orbital period, supporting our interpretation
that these lines originate in different regions (see
\S\ref{sec-discussion}). This result is consistent with the findings
from previous sections (see, e.g., Table~\ref{tab-warmabs}) whereby a
comparison between {\sc xstar} modelling of the the near-dip and
far-dip regions show that the largest changes come from the less
ionized ``warm'' region of $\log \xi \approx 3$.

\input{t6.tab}

\section{Discussion}
\label{sec-discussion} Our \chandra-HETGS spectral study of \1624\
over its orbital period confirms the detection of H- and He-like
\ion{Fe}{25} and \ion{Fe}{26} absorption. Detailed photoionization
modelling with {\sc xstar} predicts additional lines from
\ion{Ca}{20}, \ion{Ar}{17}, and \ion{S}{16}. Of particular note is not
only the confirmation of these lines with detailed photoionization
modelling, but also that the variability analysis strongly suggests
that these lines can be identified as originating from the accretion
disk corona (dominated by high ionization plasma giving rise
predominantly to \ion{Fe}{26}) and the outer accretion disk (as
indicated by the flux evolution of the lower ionization variable
\ion{Fe}{25} absorption).

While the time-averaged spectrum over the \1624 orbital period can be
modeled reasonably well with one ionization zone, variability studies
instead suggest that a model with two ionization zones is a better
description of the data. This interpretation has important
implications for geometric scenarios and our ability to map the
emitting regions as a function of orbital phase. In an earlier
analysis of this data set, \citet{iaria07} claimed velocity widths of up
to 3500~$\rm km\ s^{-1}$, to which they attribute an origin between
the external region of the ADC and the disk edge. In this
interpretation, the broadening would be due to bulk motion or
turbulence connected to the coronal activity. Our variability results
indicate that neither bulk motion nor turbulence is necessary to
produce the line widths. We arrive at this conclusion based on
variability studies where line velocity widths are seen to change over
relatively short time scales, \emph{and} that in some phases the
velocity widths are comparable to the HEG FWHM resolution (i.e., \feI
during the near-dip phase, and \feII during the far-dip phase) and
therefore are easily described by thermal broadening.

To take our two-region model further, we consider it in the context of
simple arguments relating the ionization parameter to the location of
the absorbing plasma. As pointed out by \citet{tarter69}, the
ionization parameter $\xi$ can be defined by the relation $\xi =
L_{\rm X} n_{\rm e}^{-1} R^{-2}$, where $L_{\rm X}$ is the total
un-absorbed source luminosity ($= 4.9\times10^{37}$\,ergs\,s$^{-1}$
between 1--10~keV during the epoch of our \1624 observation) and $R$
is the distance to the absorbing plasma. The electron particle density
$n_{\rm e} = N_{\rm H}/\Delta R$ is characterized by the thickness of
the absorbing medium, $\Delta R$, and the equivalent Hydrogen column of
the ionized region. Using this, we consider the possible range of $R$
for the plasma giving rise to the observed ionized absorption lines.

\begin{figure}
%\epsscale{2.0}
\plotone{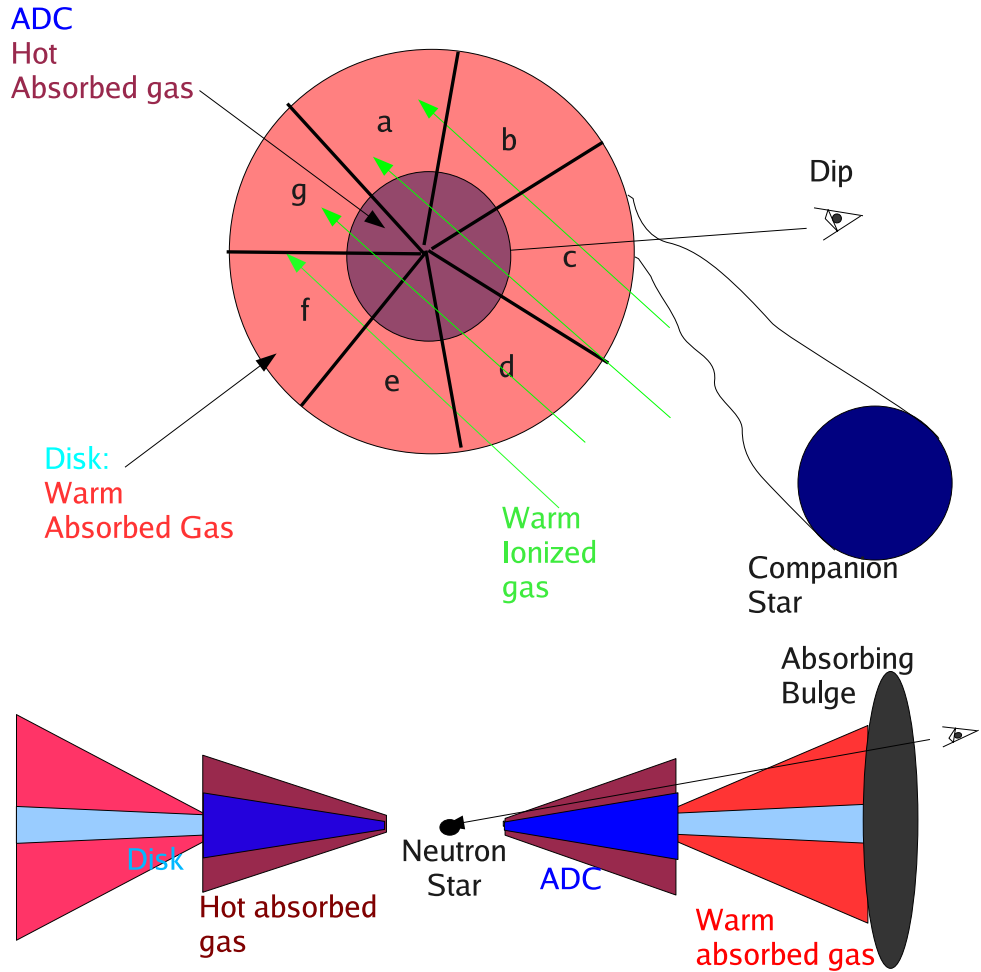}
\caption{The \1624 geometry in the context of our line-of-sight view as determined
from absorption line variability studies. 
Our studies indicate that the hotter component dominates the ADC,  whereas the 
cooler gas is associated with the outer disk rim, which intersects our 
line-of-sight. {\it a-g} refer to the \1624 orbital phase as indicated by the 
light curve of Fig.~\ref{fig-lc}a.}
\label{fig-model}
\end{figure}

A sketch of a scenario explaining the major results of such an
analysis is shown in Fig.~\ref{fig-model}. As discussed above, the
observed orbital variability of the different spectral components
points towards an extended ADC similar to the
emission geometry proposed by \citet{white88} or \citet{church04} (see
\citealt{church04b} for a summary). In this model, the ADC is large
and therefore only slight variations of its components with orbital
phase are expected, in line with our observation that the hot plasma
component is constant over the orbit. This location of the hot
component is consistent with the spectral analysis: Our modelling
efforts show that the \hot gas component can be described by $\xi_{\rm
  hot} \approx 10^{4.3}\,{\rm ergs\,cm\,s^{-1}}$ ($T \approx 3 \times
10^6$~K) ionizing a plasma of column $\approx 2 \times 10^{23} \rm
cm^{-2}$. ``Naively'' assuming a geometrically thick $\Delta R\sim R$,
a distance $R \sim3\times10^{10}$~cm is obtained, which is comparable
to the size of the Accretion Disk Corona (ADC) derived from the
relation $R_{\rm ADC} = L_{\rm X}^{0.88\pm0.16}$ of \citet{church04}.
A similar value for an extended ADC was recently determined
by \citet{schulz08b} while measuring Doppler broadening in several
broad emission lines in \cyg.

In contrast, $\xi_{\rm warm}$ changes more dramatically over the \1624
orbital phase with the largest differences seen in a comparison
between the far-dip ($\xi \approx 10^{3.4}\,{\rm ergs\,cm\,s^{-1}}$;
$T\approx 1.0\times 10^6$~K) and near-dip ($\xi \approx 10^{3.1}\,{\rm
  ergs\,cm\,s^{-1}}$; $T\approx 0.9\times 10^6$~K) phases. This
component cannot originate in the ADC: If one assumes $\Delta R\sim R$
and a column density of $N_{\rm H}\approx2\times10^{22}\rm cm^{-2}$ as
appropriate for the \warm component (Table~\ref{tab-warmabs}), the
data require $R \sim 10^{12}$~cm, which is significantly larger than
the likely accretion disk radius of $\sim10^{11}$ cm. If, however, a
geometrically thin plasma of $\Delta R/R\sim 0.1$ is assumed for this
\warm component, absorber distances between $1.8\times 10^{11}$~cm
(near-dip) and $\approx 1.1\times 10^{11}$~cm (far-dip) are
obtained. Since the radius of the accretion disk is $\approx 1.1\times
10^{11}$\,cm as obtained from the Roche Lobe size of the neutron star
\citep{church04}, it is therefore likely that the warm component
originates in the accretion disk rim, e.g., where the accretion stream
impacts onto the disk and where gas is irradiated by the neutron star.

We note that this interpretation of the absorbers is also consistent
with the variation of the broad-band spectral parameters shown in
Fig.~\ref{fig-evo}. In terms of the model of \citet{white88} and
\citet{church04}, the blackbody component is taken to originate from
the neutron star, while the power-law component is interpreted as
being due to Comptonization in the extended ADC. In this
interpretation the power-law component is expected to exhibit a
quasi-sinusoidal variation and to be weakest during the times of
dipping when the ADC is partly covered by the warm absorbing material.
A consequence of this model is also that the ratio
Fe~\textsc{xxv}/Fe~\textsc{xxvi} shows a maximum during the dipping,
since the optical depth in Fe~\textsc{xxv} is largest during that
time.  

We also note that the evolution of the spectrum can be modeled with
partial covering of a single power-law component, with the local
absorber varying over orbital phase.  In such a model, the soft-end of
the uncovered power-law replaces the blackbody component.  This offers
a slightly different explanation of the relative strength of the variation
of the absorbed power-law component, relative to the blackbody.  The
``covered power-law'' is that portion of the ADC at low latitudes,
where obscuration by the disk is expected to be greatest, and the ADC
is expected to be widest.  The ``uncovered power-law'' arises from a
geometrically narrower region at high latitudes, where obscuration by
the disk is the least.  For either the blackbody plus power-law
scenario, or the partial covering scenario, our analysis shows that
variations are predominantly driven by changes in obscuration, rather than any
intrinsic variation of the components.

\section{Summary}
\label{sec-summary}
We summarize the major conclusions and discoveries of this paper: 
\begin{itemize}
\item A re-analysis of the \chandra-HETGS spectrum of \1624 based on
  photoionization modelling and variability studies, shows that while
  the continuum spectrum can be modeled using a blackbody plus
  power-law, a slightly more preferred model is a
  single $\Gamma=2.25$ power-law partially (71\%) covered by a local
  absorber of column density $N_{\rm
    H,\ Local}=8.1_{-0.6}^{+0.7}\times 10^{22}\, \rm cm^{-2}$.  The
  total $1.6\times10^{23}\rm\, cm^{-2}$ large line-of-sight ($N_{\rm
    H,\ Local} + N_{\rm H,\ Gal}$) column therefore can attenuate the
  power-law up to 5~keV, even though the X-ray spectrum can be seen
  down to 1.7~keV. For spectra with limited (e.g. 0.5-10~keV) band
  coverage, the severe absorption at the low energies coupled with
  limited counts at the high energies can flatten a $\Gamma=2.25$
  power-law to appear as $\Gamma\sim1.2$. The source luminosity at the
  epoch of our observations is estimated to be $4.9\times 10^{37}\,\rm
  ergs\,s^{-1}$ for a distance of 15 kpc as recently
  determined by \cite{xiang07}.

\item We report the discovery of a possible
  quasi-sinusoidal modulation with period $P=43_{-9}^{+13}$~ks over
  the $\sim 76$~ks orbital period, with the caveat that we have
  observed \1624 only over one binary orbit. There is strong
  indication that variations in observed spectral properties are
  predominantly driven by changes in obscuration, rather than any
  intrinsic variation of the power-law or black-body components.

\item We confirm the detection of ionized absorption from
  \ion{Fe}{26}, \ion{Fe}{25}, and \ion{Ca}{20}, which have been
  detected by \cite{iaria07}.  {\it Detailed photoionization modeling
    with {\sc xstar} has allowed us to derive the strengths of other
    absorption lines of Mg, Si, S and Ar.}

\item Our detailed variability studies of detected strong absorption lines 
 point to a two-temperature plasma for their origin: a highly ionized 
 component of $\xi_{\rm hot} \approx 10^{4.3}\,{\rm ergs\,cm\,s^{-1}}$ 
 (T$\sim 3.0\times 10^{6}$~K) associated with an extended accretion disk 
 corona of $R \sim3\times10^{10}$~cm, and a less ionized more variable 
 component of $\xi \approx 10^{3.4}\,{\rm ergs\,cm\,s^{-1}}$ 
 (T$\sim 1.0\times 10^{6}$~K; far-dip) and 
 $\rm \xi \approx 10^{3.1}\, ergs\,cm\,s^{-1}$ 
 (T$\sim 9.0\times 10^{5}$~K; near-dip) coincident with the accretion disk 
 rim. This model is offered in favor of bulk Comptonization, as previously
 suggested, as an explanation for observed line broadening.

\item We also report \ion{Fe}{25} and \ion{Fe}{26} absorption line 
 variations (in wavelength, strength, width) over the \1624 orbital period.
 This, with previously noted variations have allowed us to map plasma 
 properties to our viewing of \1624 over its $\sim 76$~ks orbital period 
 (see Fig.~\ref{fig-model}). 

\end{itemize}

\acknowledgments{This work was funded
  by the NASA / \chandra\ grant GO4-3056X; we are thankful for its
  support. JCL thanks the Harvard University Faculty of Arts and
  Sciences for supplemental financial support.}

\end{document}

%% file: t1.tab
\begin{deluxetable}{lclllll}
\tabletypesize{\scriptsize}
\tablecaption{Best-fit continuum and broad Fe emission line parameters}
\tablecolumns{7}
\tablehead{
\colhead{Component} &
\colhead{Parameter} &
\colhead{persistent}  & 
\colhead{pre-dip} &
\colhead{post-dip} &
\colhead{near-dip} &
\colhead{far-dip} }
\startdata
 & $\chi^{2}/d.o.f$ & $1435/1338$ & $1003/992$ & $1205/1162$ & $1026/1066$ & $1248/1194$\\
 & $L_X$ (erg s$^{-1}$) & $4.9\times 10^{37}$ & $4.8\times 10^{37}$ & $5.0\times 10^{37}$ & $4.6\times 10^{37}$ & $4.9\times 10^{37}$\\
{\sc tbabs}$^a$ & $N_H$ (10$^{22}$ cm$^{-2}$) & $7.6_{-0.4}^{+0.4}$ & $7.4_{-0.4}^{+0.6}$ & $7.8_{-0.5}^{+0.5}$ & $7.5_{-0.5}^{+0.6}$ & $7.6_{-0.4}^{+0.4}$\\
Blackbody & kT (keV) & $1.39_{-0.04}^{+0.05}$ & $1.38_{-0.05}^{+0.08}$ & $1.43_{-0.05}^{+0.06}$ & $1.43_{-0.06}^{+0.08}$ & $1.35_{-0.04}^{+0.06}$\\
 & Norm (10$^{-2}$) & $1.42_{-0.11}^{+0.15}$ & $1.52_{-0.29}^{+0.46}$ & $1.53_{-0.12}^{+0.14}$ & $1.45_{-0.16}^{+0.27}$ & $1.41_{-0.15}^{+0.24}$\\
Powerlaw & $\Gamma$ & $1.14_{-0.48}^{+0.33}$ & $0.65_{-1.70}^{+0.64}$ & $1.43_{-0.55}^{+0.43}$ & $1.13_{-1.17}^{+0.54}$ & $1.06_{-0.71}^{+0.39}$\\
  & Norm ($10^{-2}$\ ph\ cm$^{-2}$\ s$^{-1}$\ keV$^{-1}$) & $6.2_{-4.1}^{+5.8}$ & $2.7_{-2.6}^{+8.3}$ & $9.1_{-6.4}^{+9.9}$ & $5.1_{-4.7}^{+8.8}$ & $5.8_{-4.7}^{+7.2}$\\
Fe emission & $E_{\rm Fe}$ (keV) & $6.76_{-0.17}^{+0.20}$ & $6.57_{-0.19}^{+0.19}$ & $6.66_{-0.08}^{+0.30}$ & $6.97_{-0.31}^{+0.47}$ & $6.68_{-0.07}^{+0.07}$\\
feature &  & (1.83 \AA) & (1.89\AA) &(1.86\AA) &(1.78\AA) &(1.86\AA)   \\
 & Flux $I_{\rm Fe}$ (10$^{-3}$\ ph\ cm$^{-2}$\ s$^{-1}$) & $0.9_{-0.3}^{+0.4}$ & $0.8_{-0.4}^{+0.5}$ & $0.6_{-0.3}^{+0.4}$ & $1.2_{-0.6}^{+0.6}$ & $0.8_{-0.2}^{+0.2}$\\
  & Width $\sigma$ (keV) & $0.34_{-0.14}^{+0.19}$ & $0.24_{-0.12}^{+0.19}$ & $0.13_{-0.07}^{+0.30}$ & $0.50_{-0.15}^{+0.32}$ & $0.18_{-0.05}^{+0.07}$\\
  & ($10^3\ \rm km \ s^{-1}$) & ($15_{-6}^{+8}$)  & ($11_{-6}^{+9}$) & ($6_{-3}^{+14}$) & ($22_{-7}^{+14}$) & ($8_{-2}^{+3}$) \\ 
\enddata
\tablecomments{$^a${\sc tbabs} is the absorption model described in \S\ref{sec-broad}. Errors are quoted at 90\% confidence.}
\label{tab-cont}
\end{deluxetable}

%% file: t2.tab
\begin{deluxetable}{lllrrrccl}
\tabletypesize{\scriptsize}
\tablecaption{Ions predicted from {\sc xstar} fit to time averaged spectrum }\tablecolumns{9}
\tablehead{\colhead{Ion$^a$} & \colhead{Transition$^b$} & \colhead{$f_{ij}^c$} & \colhead{$\lambda_{lab}^d$} & \colhead{$\lambda_{obs}^e$} & \colhead{$\tau^f$} & \colhead{$W_{\lambda}^g$} & \colhead{model$^h$} & \colhead{comments}
}
\startdata
Mg {\sc xii} & Ly $\beta$:  $1s(^2S_{1/2})\rightarrow 3p(^2P^o)$ & 0.079 & 7.106 & 7.111 & 6.75 & 0.51 & w2 & weak\\
Si {\sc xiv} & Ly $\alpha$:  $1s(^2S_{1/2})\rightarrow 2p(^2P^o)$ & 0.414 & 6.182 & 6.170 & 5.45 & 1.10 & w1 &weak\\
 & & & & 6.186 & 81.70 & 4.10 & w2 & strong\\
Si {\sc xiv} & Ly $\beta$:  $1s(^2S_{1/2})\rightarrow 3p(^2P^o)$ & 0.079 & 5.217 &  5.221 & 13.70 & 0.70 & w2 & weak\\
S {\sc xvi} & Ly $\alpha$:  $1s(^2S_{1/2})\rightarrow 2p(^2P^o)$ & 0.413 & 4.729 & 4.720 & 6.31 & 0.98 & w1 &weak\\
 & & & & 4.732 & 83.00 & 3.20 & w2 & strong\\
S {\sc xvi} & Ly $\beta$:  $1s(^2S_{1/2})\rightarrow 3p(^2P^o)$ & 0.079 & 3.991 & 3.994 & 14.20 & 0.58 & w2 & weak\\
Ar {\sc xviii} & Ly $\alpha$:  $1s(^2S_{1/2})\rightarrow 2p(^2P^o)$ & 0.412 & 3.733 & 3.736 & 22.50 & 0.79 & w2 & weak\\
Ca {\sc xx} & Ly $\alpha$:  $1s(^2S_{1/2})\rightarrow 2p(^2P^o)$ & 0.411 & 3.020 & 3.022 & 12.00 & 0.62 & w2 & weak\\

Fe {\sc xxv} & He $\alpha$:  $1s^2(^1S_{0})\rightarrow 1s2p(^1P_{1}^o)$ & 0.775 & 1.850 &  1.851 & 113.00 & 2.30 & w2 & strong\\
Fe {\sc xxv} & He $\beta$:  $1s^2(^1S_{0})\rightarrow 1s3p(^1P_{1}^o)$ & 0.152 & 1.573 & 1.574 & 31.00 & 0.74 & w2 &weak\\
Fe {\sc xxvi} & Ly $\alpha$:  $1s(^2S_{1/2})\rightarrow 2p(^2P^o)$ & 0.408 & 1.780 & 1.776 & 103.00 & 4.00 & w1 &strong\\
 & & & & 1.781 & 145.00 & 2.10 & w2 & strong\\
Fe {\sc xxvi} & Ly $\beta$:  $1s(^2S_{1/2})\rightarrow 3p(^2P^o)$ & 0.078 & 1.503 & 1.500 & 12.70 & 0.72 & w1 &weak\\
 & & & & 1.504 & 29.60 & 0.65 & w2 & weak\\
Fe {\sc xxvi} & Ly $\gamma$:  $1s(^2S_{1/2})\rightarrow 4p(^2P^o)$ & 0.029 & 1.438 & 1.435 & 5.14 & 0.26 & w1 & weak\\
 & & & & 1.439 & 10.90 & 0.26 & w2 & weak\\
Fe {\sc xxvi} & Ly $\delta$:  $1s(^2S_{1/2})\rightarrow 5p(^2P^o)$ & 0.014 & 1.404 & 1.401 & 2.65 & 0.12 & w1 & weak\\
 & & & & 1.405 & 3.77 & 0.13 & w2 & weak\\
\enddata
\tablecomments{ $^c$Oscillator Strength, $^d$Laboratary Wavelength, $^e$Observed wavelength, $^f$Optical Depth, $^g$Equivalent Width, $^h$``w1'' and ``w2'' refer to the absorber-1 and  absorber-2 respectively.}
\label{tab:xstar}
\end{deluxetable}

%% file: t3.tab
\begin{deluxetable}{ccccccccc}
\tabletypesize{\scriptsize}
\tablecaption{Best-fit \feI and \feII absorption lines az measured in the different phases.}
\tablehead{ \colhead{Orbital} &
\colhead{Wavelength} &
\colhead{Width} &
\colhead{EW} &
\colhead{Line} &
\colhead{Wavelength} &
\colhead{Width} &
\colhead{EW} &
\colhead{Line} \\
\colhead{phase} &
\colhead{(\AA)(keV)} &
\colhead{(m\AA)(km/s)} &
\colhead{(m\AA)} &
\colhead{Flux} &
\colhead{(\AA)(keV)} &
\colhead{(m\AA)(km/s)} &
\colhead{(m\AA)} &
\colhead{Flux}}
\startdata
& \multicolumn{4}{c}{Fe {\sc XXVI}} & \multicolumn{4}{c}{Fe {\sc XXV}} \\\\ 
nondip & $1.778_{-0.002}^{+0.001}(6.97)$ & $ 5_{- 2}^{+ 2}(770_{-358}^{+383})$ & $7.2_{-1.5}^{+1.3}$ & $4.4_{-0.9}^{+0.8}$ & $1.849_{-0.004}^{+0.007}(6.71)$ & $<16.4(<2659)$ & $3.1_{-1.4}^{+2.3}$ & $1.9_{-0.8}^{+1.4}$\\
predip & $1.780_{-0.003}^{+0.003}(6.97)$ & $<7.0(<1176)$ & $5.5_{-2.0}^{+2.2}$ & $3.6_{-1.3}^{+1.4}$ & $1.848_{-0.003}^{+0.004}(6.71)$ & $<10.0(<1623)$ & $4.6_{-2.0}^{+2.3}$ & $3.0_{-1.3}^{+1.5}$\\
postdip & $1.775_{-0.002}^{+0.003}(6.99)$ & $ 4_{- 3}^{+ 3}(758_{-495}^{+439})$ & $6.0_{-1.6}^{+1.6}$ & $3.7_{-1.0}^{+1.0}$ & $1.858_{-0.012}^{+0.008}(6.67)$ & $12_{- 8}^{+ 8}(1934_{-1222}^{+1295})$ & $4.5_{-2.2}^{+2.7}$ & $2.8_{-1.3}^{+1.7}$\\
neardip & $1.778_{-0.001}^{+0.002}(6.97)$ & $<3.8(<635)$ & $5.1_{-1.3}^{+1.4}$ & $3.1_{-0.8}^{+0.9}$ & $1.857_{-0.004}^{+0.003}(6.68)$ & $ 7_{- 4}^{+ 6}(1171_{-571}^{+1049})$ & $6.8_{-2.1}^{+2.6}$ & $4.1_{-1.3}^{+1.6}$\\
fardip & $1.778_{-0.003}^{+0.002}(6.97)$ & $ 6_{- 3}^{+ 3}(1094_{-427}^{+492})$ & $8.2_{-1.8}^{+1.9}$ & $5.0_{-1.1}^{+1.2}$ & $1.847_{-0.002}^{+0.001}(6.71)$ & $<4.6(<746)$ & $3.0_{-1.2}^{+1.1}$ & $1.9_{-0.7}^{+0.7}$\\
\enddata
\tablecomments{EW is the equivalent width, and the line flux is in units of $10^{-4}$ ph cm$^{-2}$ s$^{-1}$. Errors are quoted at 90\% confidence.}
\label{tab-gauss}
\end{deluxetable}

%% file: t4.tab
\begin{deluxetable}{ccccccc}
\tabletypesize{\scriptsize}
\tablecaption{Best-fit parameters from {\sc xstar} photoionization models of the near-dip and far-dip phases.}
\tablehead{ \colhead{} &
\colhead{$\log(\xi)$} &
\colhead{$\log(N_{\rm H})$} &
\colhead{$v$} &
\colhead{$\log(\xi)$} &
\colhead{$\log(N_{\rm H})$} &
\colhead{$v$}\\
\colhead{phase} &
\colhead{(ergs~cm~s$^{-1}$)} &
\colhead{(cm$^{-2}$)} &
\colhead{(km/s)} &
\colhead{(ergs~cm~s$^{-1}$)} &
\colhead{(cm$^{-2}$)} &
\colhead{(km/s)}}
\startdata
 & \multicolumn{3}{c}{Hot Absorption} & \multicolumn{3}{c}{Warm Absorption}\\
 near-dip &    $4.2_{-0.6}^{+0.8}$ &     $23.4_{-0.6}^{+0.5}$ &     $335_{-318}^{+346}$ &     $3.1_{-0.2}^{+0.2}$ &     $22.3_{-0.4}^{+0.3}$ &     $-276_{-172}^{+271}$\\
 far-dip &     $4.3_{-0.6}^{+0.7}$ &     $23.3_{-0.3}^{+0.3}$ &     $-263_{-310}^{+306}$ &     $3.4_{-0.1}^{+0.2}$ &     $22.2_{-0.2}^{+0.2}$ &   $1028_{-307}^{+4890}$ \\
\enddata
\tablecomments{$\xi$ is the photoionization parameter from {\sc xstar}. ``$N_{\rm H}$'' is the equivalent Hydrogen column density in the associated ionized region. ``$v$'' is the measured velocity shift, (-) blueshift towards us and (+) redshift away us. Errors are quoted at 90\% confidence.}
\label{tab-warmabs}
\end{deluxetable}

%% file: t5.tab
\begin{deluxetable}{ccccccccccc}
\tabletypesize{\scriptsize}
\tablecaption{Best-fit parameters over $\sim$10~ks time step.}
\tablehead{ \colhead{} &
 \colhead{$N_{\rm H}$} &
\colhead{kT} &
 \colhead{uFlux} &
\colhead{Flux} &
 \colhead{$\Gamma$ } &
\colhead{uFlux} &
 \colhead{Flux} &
\colhead{$E$} & 
 \colhead{Width} &
\colhead{Flux} \\
\colhead{phase} &
 \colhead{($10^{22}\,\rm cm^{-2}$)} &
\colhead{(keV)} &
 \colhead{} &
\colhead{} &
 \colhead{} &
\colhead{} &
\colhead{} & 
 \colhead{(KeV)} &
 \colhead{(KeV)} &
\colhead{} \\ 
\colhead{} & 
 \colhead{} & 
 \colhead{} & 
\colhead{} & 
 \colhead{} & 
 \colhead{} & 
\colhead{} & 
 \colhead{} & 
 \colhead{(\AA)} & 
\colhead{($10^3$ km/s)} & 
 \colhead{}} 
\startdata
& & \multicolumn{3}{c}{blackbody} & \multicolumn{3}{c}{power-law} & \multicolumn{3}{c}{Fe emission feature}\\
a & $7.6_{-0.4}^{+0.7}$ & $1.35_{-0.05}^{+0.09}$ & $11.7_{-2.2}^{+1.1}$ & $6.4_{-1.3}^{+0.5}$ & $0.8_{-0.8}^{+0.7}$ & $7.7_{-3.4}^{+4.5}$ & $4.9_{-0.8}^{+1.3}$ & $6.67_{-0.08}^{+0.12}$ & $0.25_{-0.11}^{+0.08}$ & $1.7_{-0.5}^{+0.5}$\\ & & & & & & & & $(1.858)$ & $(11_{-5}^{+4})$ & \\
b & $7.3_{-0.3}^{+1.0}$ & $1.32_{-0.06}^{+0.07}$ & $11.3_{-3.0}^{+1.6}$ & $6.2_{-1.8}^{+0.9}$ & $0.5_{-0.5}^{+0.9}$ & $6.5_{-2.0}^{+5.7}$ & $4.5_{-1.1}^{+2.0}$ & $6.44_{-0.25}^{+0.41}$ & $<0.8$ & $0.4_{-0.4}^{+0.5}$\\ & & & & & & & & $(1.925)$ & ($<37$) & \\
c & $7.4_{-0.3}^{+0.4}$ & $1.46_{-0.11}^{+0.12}$ & $12.0_{-2.2}^{+2.4}$ & $6.8_{-1.4}^{+1.3}$ & $1.0_{-1.0}^{+1.3}$ & $4.1_{-1.6}^{+4.6}$ & $2.5_{-1.6}^{+1.6}$ & $6.96_{-0.28}^{+1.01}$ & $0.19_{-0.22}^{+0.60}$ & $0.6_{-0.5}^{+0.7}$\\ & & & & & & & & $(1.781)$ & $(49_{-46}^{+151})$ & \\
d & $8.0_{-0.9}^{+1.4}$ & $1.48_{-0.11}^{+0.12}$ & $12.6_{-2.2}^{+1.0}$ & $6.9_{-1.3}^{+0.5}$ & $1.7_{-1.4}^{+1.1}$ & $8.6_{-5.9}^{+13.9}$ & $3.5_{-0.8}^{+1.5}$ & $6.91_{-0.16}^{+0.19}$ & $0.30_{-0.14}^{+0.19}$ & $1.6_{-0.7}^{+0.8}$\\ & & & & & & & & $(1.794)$ & $(13_{-6}^{+8})$ & \\
e & $7.4_{-0.3}^{+0.6}$ & $1.32_{-0.06}^{+0.08}$ & $11.2_{-1.8}^{+2.9}$ & $6.1_{-1.5}^{+0.7}$ & $0.7_{-0.7}^{+0.7}$ & $7.8_{-2.9}^{+5.0}$ & $5.2_{-0.8}^{+1.6}$ & $6.58_{-0.19}^{+0.18}$ & $0.21_{-0.11}^{+0.25}$ & $1.0_{-0.7}^{+0.7}$\\ & & & & & & & & $(1.884)$ & $(10_{-5}^{+11})$ & \\
f & $7.6_{-0.6}^{+1.2}$ & $1.31_{-0.05}^{+0.14}$ & $11.7_{-1.6}^{+1.3}$ & $6.3_{-0.9}^{+0.7}$ & $1.1_{-1.5}^{+1.1}$ & $6.8_{-5.1}^{+7.4}$ & $3.8_{-0.8}^{+1.1}$ & $6.69_{-0.22}^{+0.23}$ & $0.28_{-0.17}^{+0.41}$ & $1.0_{-0.5}^{+0.6}$\\ & & & & & & & & $(1.853)$ & $(13_{-8}^{+18})$ & \\
g & $7.7_{-0.5}^{+0.7}$ & $1.26_{-0.05}^{+0.06}$ & $11.8_{-1.6}^{+1.7}$ & $6.2_{-1.0}^{+0.9}$ & $1.1_{-1.1}^{+0.8}$ & $5.5_{-3.1}^{+4.5}$ & $3.2_{-1.1}^{+1.0}$ & $6.82_{-0.14}^{+0.34}$ & $0.26_{-0.12}^{+0.29}$ & $1.4_{-0.6}^{+1.3}$\\ & & & & & & & & $(1.818)$ & $(11_{-5}^{+12})$ & \\
\enddata
\tablecomments{$N_{\rm H}$ is the LOS cold Hydrogen column density. uFlux and Flux refer respectively to the unabsorbed and absorbed integrated flux [$10^{-10}$ ergs cm$^{-2}$ s$^{-1}$] over 0.6-10.0~keV. $I_{\rm line}$ is the flux [$10^{-2}$~ph~cm$^{-2}$~s$^{-1}$] of the broad Fe emission line. Errors are quoted at 90\% confidence.}
\label{tab-7con}
\end{deluxetable}

%% file: t6.tab
\begin{deluxetable}{ccccccccc}
\tabletypesize{\scriptsize}
\tablecaption{Best-Fit parameters of Fe H-like and He-like absorption lines.}
\tablehead{ \colhead{} &
 \colhead{Wavelength} &
\colhead{Width} &
\colhead{EW} &
 \colhead{Flux} &
\colhead{Wavelength} &
 \colhead{Width} & 
\colhead{EW} &
 \colhead{Flux}\\
\colhead{phase} &
 \colhead{(\AA)(keV)} &
\colhead{(m\AA)(km/s)} &
 \colhead{(m\AA)} &
\colhead{} &
\colhead{(\AA)(keV)} &
 \colhead{(m\AA)(km/s)} &
\colhead{(m\AA)} &
\colhead{}}
\startdata
& \multicolumn{4}{c}{Fe {\sc XXVI}} & \multicolumn{4}{c}{Fe {\sc XXV}} \\\\a & $1.780_{-0.002}^{+0.003}(6.97)$ & $<7.4(<1250)$ & $5.9_{-2.0}^{+1.7}$ & $4.1_{-1.4}^{+1.2}$ & $1.845_{-0.002}^{+0.003}(6.72)$ & $<6.2(<1006)$ & $4.7_{-2.1}^{+1.8}$ & $3.2_{-1.4}^{+1.2}$\\
b & $1.780_{-0.003}^{+0.003}(6.97)$ & $<6.2(<1042)$ & $5.0_{-2.2}^{+1.8}$ & $3.2_{-1.4}^{+1.2}$ & $1.852_{-0.006}^{+0.010}(6.69)$ & $<20.6(<3342)$ & $6.0_{-3.4}^{+4.4}$ & $3.8_{-2.2}^{+2.8}$\\
c & $1.782_{-0.002}^{+0.001}(6.96)$ & $<4.4(<749)$ & $7.8_{-2.5}^{+1.9}$ & $4.2_{-1.3}^{+1.0}$ & $1.850_{-0.010}^{+0.007}(6.70)$ & $13_{- 7}^{+ 7}(2077_{-1061}^{+1166})$ & $13.1_{-5.6}^{+7.0}$ & $7.1_{-3.0}^{+3.8}$\\
d & $1.775_{-0.002}^{+0.003}(6.99)$ & $<3.9(<667)$ & $5.9_{-2.2}^{+1.8}$ & $3.9_{-1.4}^{+1.2}$ & $1.860_{-0.004}^{+0.000}(6.67)$ & $<6.9(<1110)$ & $5.7_{-2.2}^{+1.9}$ & $3.8_{-1.4}^{+1.2}$\\
e & $1.776_{-0.005}^{+0.004}(6.98)$ & $ 8_{- 4}^{+ 5}(1377_{-648}^{+878})$ & $10.4_{-3.2}^{+3.9}$ & $6.7_{-2.1}^{+2.5}$ & $1.845_{-0.005}^{+0.015}(6.72)$ & $<10.0(<1626)$ & $<6.5$ & $<4.2$\\
f & $1.777_{-0.007}^{+0.010}(6.98)$ & $<15.0(<2532)$ & $11.6_{-4.6}^{+4.2}$ & $7.1_{-2.8}^{+2.5}$ & $1.845_{-0.005}^{+0.003}(6.72)$ & $<4.9(<803)$ & $3.3_{-2.4}^{+2.2}$ & $2.0_{-1.5}^{+1.3}$\\
g & $1.780_{-0.005}^{+0.006}(6.97)$ & $9.6_{-5.3}^{+5.7}(2028_{-1112}^{+1161})$ & $14.5_{-5.3}^{+5.5}$ & $8.3_{-3.0}^{+3.1}$ & $1.850_{-0.005}^{+0.003}(6.70)$ & $<7.5(<1218)$ & $3.8_{-2.7}^{+2.2}$ & $2.2_{-1.5}^{+1.3}$\\
\enddata
\tablecomments{EW is the equivalent width and the unit of flux is $10^{-4}$ ph cm$^{-2}$ s$^{-1}$. Errors are quoted at 90\% confidence.}
\label{tab-Fe}
\end{deluxetable}